\documentclass[twocolumn]{svjour3}          % twocolumn
\smartqed  % flush right qed marks, e.g. at end of proof
\usepackage{graphicx}
\usepackage{cite}

\journalname{J Supercond Nov Magn}
\begin{document}

\title{High-T$_c$ and Low-T$_c$ Superconductivity in Electron Systems With Repulsion %\thanks{Grants or other notes
} %\subtitle{Do you have a subtitle? \\ If so, write it here}


\author{Maxim Yu. Kagan$^{1,2}$         \and
        Vitaly A. Mitskan$^{3,4}$ \and
        Maxim M. Korovushkin$^{3,5}$ %etc.
}


\institute{$^1$P.\,L. Kapitza Institute for Physical Problems,
119334 Moscow, Russia\\ $^2$National Research
University Higher School of Economics,\\
109028 Moscow, Russia\\
$^3$L.\,V. Kirensky Institute of Physics,
660036 Krasnoyarsk, Russia\\
$^4$Siberian State Aerospace University, 660014 Krasnoyarsk, Russia\\
$^5$Nordita, KTH Royal Institute of Technology and Stockholm
University, Roslagstullsbacken 23, SE-106 91 Stockholm, Sweden\\}
%\email{kagan@kapitza.ras.ru}  }

\date{Received: date / Accepted: date}

\maketitle

\begin{abstract}
We demonstrate the instability of the normal state of purely
repulsive fermionic systems towards the transition to the
Kohn-Luttinger superconducting state. We construct the
superconducting phase diagrams of these systems in the framework
of the Hubbard and Shubin-Vonsovsky models on the square and
hexagonal lattices. We show that an account for the long-range
Coulomb interactions, as well as the Kohn- Luttinger
renormalizations lead to an increase in the critical
superconducting temperatures in various materials, such as
high-temperature superconductors, idealized monolayer and bilayer
of doped graphene. Additionally, we discuss the role of the
structural disorder and the nonmagnetic impurities in
superconducting properties of real graphene systems.
\end{abstract}

\section{Introduction}
\label{intro}

In recent years, a significant progress in experimental and
theoretical investigation of high-temperature and low-temperature
superconducting systems with nonphonon nature of the Cooper
pairing and nontrivial structure of the order parameter has been
achieved. Along with the numerous studies of superconducting
properties of these systems using pairing mechanisms caused by
electron correlations and other exotic superconductivity
mechanisms, some authors widely discuss the possibility of the
development of Cooper instability in new superconducting systems
using the Kohn-Luttinger mechanism, suggesting the transformation
of initial repulsive interaction of two particles in vacuum into
the effective attraction in the presence of the fermionic
background~\cite{Kohn65,Fay68,Kagan88,Baranov92b,Kagan15b}. In
this paper, we consider the fermionic systems, such as
high-temperature superconductors and monolayer and bilayer of
doped graphene, in which the anomalous $p$-, $d$-, and $f$-wave
pairing are realized, and show that in many cases the proposed
mechanism results in quite high superconducting transition
temperatures.
%
% Moreover, for
%electron concentrations close to the Van Hove singularity (VHS) in
%the electron density of states (DOS), the superconducting
%transition temperatures increase still further and may reach the
%values of the order of $100\,K$ even in the one-band case for
%intermediate values of the ratio of the Hubbard repulsion
%parameter to the conduction band width ($U/W$).

\section{The Kohn-Luttinger Superconductivity in the Hubbard and Shubin-Vonsovsky Models}
\label{sec2}

The Hubbard model~\cite{Hubbard63} with the Hamiltonian
\begin{eqnarray}\label{Hubbard_momentum}
\hat{H} &=&\sum\limits_{\vec{p}\sigma}(\varepsilon_{\vec{p}}-\mu)
c^{\dagger}_{\vec{p}\sigma}c_{\vec{p}\sigma} +
U\sum_{\vec{p}\vec{p'}\vec{q}}c^{\dagger}_{\vec{p}\uparrow}
c^{\dagger}_{\vec{p'}+\vec{q}{\downarrow}}c_{\vec{p}+\vec{q}{\downarrow}}
c_{\vec{p'}{\uparrow}}
\end{eqnarray}
is the minimal model taking into account the band motion of
electrons in a solid and strong electron interaction. Since the
end of 1980s, a lot of experimental data on cuprates indicated
that the main dynamics of Fermi excitations evolves in the CuO$_2$
planes and that is why the 2D Hubbard model on a simple square
lattice was mainly used to describe the nonphonon mechanisms of
high-T$_c$ superconductivity. Figure 1 depicts the superconducting
phase diagram of the Hubbard model. In the region of low electron
densities $0<n<0.52$, superconductivity with the $d_{xy}$-wave
symmetry of the order parameter is
realized~\cite{Baranov92b,Baranov92a}. In the interval
$0.52<n<0.58$, the strong competition between $p$-wave and
$d_{xy}$-wave pairings takes
place~\cite{Hlubina99,Raghu12,Kagan13}. For $n
> 0.58$, superconductivity of the $d_{x^2-y^2}$-wave type
dominates~\cite{Raghu10}. It should be noted that the maximal
$T_c$ in the 2D Hubbard model was obtained in~\cite{Raghu10} in
the regime $U/W\sim1$ ($W$ is the bandwidth) for optimal electron
concentrations $n\sim0.8-0.9$. According to the
estimation~\cite{Raghu10}, the superconducting transition
temperature can reach desirable values
$T^{d_{x^2-y^2}}_c\approx100\,\textrm{K}$, which are quite
reasonable for optimally doped cuprates.

The important question concerning the role of the long-range part
of Coulomb interaction in nonphonon superconductivity was
considered in the paper~\cite{Alexandrov11}. The authors noted
that previous investigations of the Kohn-Luttinger
superconductivity were limited to the inclusion of the only
short-range Coulomb interaction $U$ having in mind the
computational difficulties connected with taking into account the
Fourier transform of the long-range Coulomb repulsion
$V_{\vec{q}}$ in the first- and second-order diagrams for the
effective interaction~\cite{Kohn65}. The
authors~\cite{Alexandrov11} choose the long-range Coulomb
interaction $V_{\vec{q}}$ in the form of the Fourier transform of
the Yukawa potential which has the standard form in the 3D case:
\begin{equation}\label{screening}
V_{\vec{q}}=\frac{4\pi e^2}{q^2+\kappa^2},
\end{equation}
where $\kappa$ is the inverse screening length. It was concluded
in~\cite{Alexandrov11} that small and intermediate values of $U$
in the presence of the long-range part of Coulomb interaction do
not induce realization of the Cooper instability in 3D and 2D
Fermi systems in the $p$-wave and $d$-wave channels, irrespective
of the value of the screening length.
\begin{figure}
\begin{center}
  \includegraphics[width=0.42\textwidth]{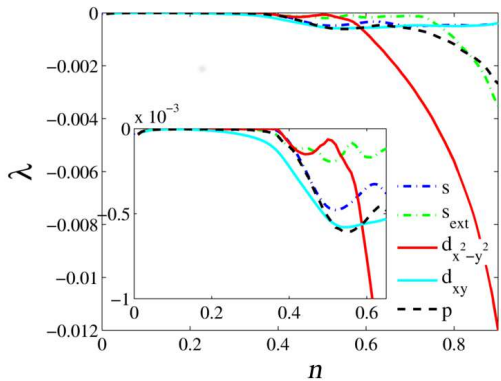}
\caption{Pairing strengths for the 2D Hubbard model at $t_2=0$ as
a function of electron concentration.}
\label{DOS}       % Give a unique label
\end{center}
\end{figure}

To clarify the role of the long-range Coulomb interaction in the
implementation of unconventional superconductivity, the
authors~\cite{Kagan11} analyzed the conditions for the occurrence
of the Kohn-Luttinger pairing in the 3D and 2D Shubin-Vonsovsky
model (extended Hubbard model) \cite{Shubin34} with Coulomb
repulsion of electrons at the neighboring sites on the square
lattice. In the momentum representation, the Hamiltonian of the
model has the form
\begin{eqnarray}\label{SVHam}
\hat{H} &=&\sum\limits_{\vec{p}\sigma}(\varepsilon_{\vec{p}}-\mu)
c^{\dagger}_{\vec{p}\sigma}c_{\vec{p}\sigma} +
U\sum_{\vec{p}\vec{p'}\vec{q}}c^{\dagger}_{\vec{p}\uparrow}
c^{\dagger}_{\vec{p'}+\vec{q}{\downarrow}}c_{\vec{p}+\vec{q}{\downarrow}}
c_{\vec{p'}{\uparrow}}\nonumber\\
&+&\frac12\sum_{\vec{p}\vec{p'}\vec{q}\sigma\sigma'}V_{\vec{p}-\vec{p'}}\,c^{\dagger}_{\vec{p}\sigma}
c^{\dagger}_{\vec{p'}+\vec{q}{\sigma'}}c_{\vec{p}+\vec{q}{\sigma'}}c_{\vec{p'}{\sigma}},
\end{eqnarray}
where the Fourier transform of the Coulomb interaction of
electrons at the nearest sites ($V_1$) and at the next-nearest
sites ($V_2$) in the 2D case on the square lattice yields
\begin{equation}\label{Vq}
V_{\vec{q}}=2V_1(\textrm{cos}\,q_x+\textrm{cos}\,q_y)+4V_2
\textrm{cos}\,q_x~\textrm{cos}\,q_y.
\end{equation}
The last term in the Hamiltonian (\ref{SVHam}) reflects the fact
that the screening radius in the systems may be by several times
larger than the unit cell parameter. It demonstrates an advantage
of the Shubin-Vonsovsky model, in which the intersite Coulomb
interaction is taken into account within several coordination
spheres.

In~\cite{Kagan11}, instead of Yukawa potential used as the Fourier
transform of the intersite interaction, the case of extremely
strong Coulomb repulsion was considered. In this case, the
Shubin-Vonsovsky model becomes the most repulsive and the most
unbeneficial model for superconductivity. However, the previous
results~\cite{Kagan88,Baranov92b} for the Kohn-Luttinger
superconducting $p$-wave pairing being attained both in the 2D and
3D Hubbard model and the same expressions for $T_c$, as in the
case of $V_1=0$, were obtained~\cite{Kagan11}. An account for
$V_1$ changes only the preexponential factor. Therefore,
superconducting $p$-wave pairing can be always realized in the
Fermi systems with pure Coulomb repulsion in the absence of
electron-phonon interaction.

A similar analysis was carried out by the authors~\cite{Raghu12}
for the extended Hubbard model in the Born weak-coupling
approximation ($W>U>V_1$). In the calculation~\cite{Raghu12} of
the effective interaction of electrons, the intersite Coulomb
interaction was taken into account only in the first order of
perturbation theory in the form (\ref{Vq}) and the polarization
contributions included only the terms of the order $U^2$. It was
shown that the long-range interaction has a tendency to suppress
unconventional pairing in some channels; nevertheless the
Kohn-Luttinger superconductivity survives in whole range of
electron densities $0<n<1$ and for all the relations between the
model parameters.

\begin{figure}
\begin{center}
  \includegraphics[width=0.38\textwidth]{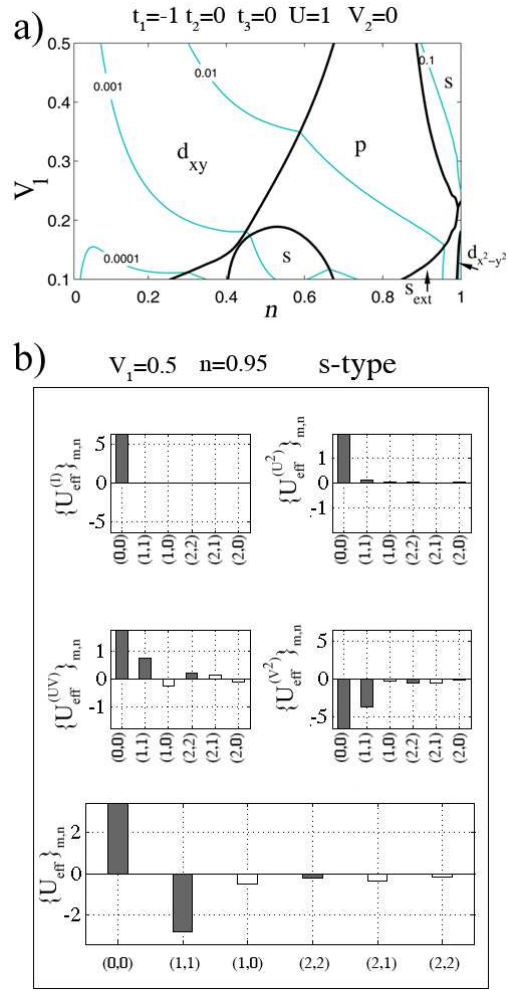}
\caption{Phase diagram of the Shubin-Vonsovsky model, constructed
taking into account the second-order contributions in $V$ for the
set of parameters $t_2=t_3=0,~U=1$ (blue curves show the lines of
constant values of $|\lambda|$) (a) and the values of the matrix
element for partial contributions $U_{eff}^{(I)}$,
$U^{(U^2)}_{eff}$, $U^{(UV)}_{eff}$, $U^{(V^2)}_{eff}$ as well as
the resulting effective interaction $\{U_{eff}\}_{mn}$ for
$V_1=0.5$ and $n=0.95$ (b).}
\label{histogram}       % Give a unique label
\end{center}
\end{figure}

The results of~\cite{Raghu12} suggest to study the conditions for
the appearance of the Kohn-Luttinger instability taking into
account all the second-order terms in the long-range Coulomb
interaction. In~\cite{Kagan13}, we considered the effect of the
Coulomb interaction of electrons $V_1$ and $V_2$ on the
realization of the Cooper instability in the framework of the
Shubin-Vonsovsky model in the Born weak-coupling approximation.
Since the polarization effects are manifested through the
second-order contributions in $V$, to account for the
Kohn-Luttinger effects connected with the intersite Coulomb
repulsion, we used the complete expression for the effective
interaction
\begin{eqnarray}\label{Gamma_wave_1}
U_{eff}(\vec{p},\vec{q})&=&U_{eff}^{(I)}(\vec{p},\vec{q})+
U_{eff}^{(U^2)}(\vec{p},\vec{q})\nonumber\\
&+&U_{eff}^{(UV)}(\vec{p},\vec{q})+U_{eff}^{(V^2)}(\vec{p},\vec{q}).
\end{eqnarray}
In this case, the polarization effects proportional to $UV$ and
$V^2$ considerably modify and complicate the structure of the
superconducting phase diagram (Fig.~\ref{histogram}a). With the
increase of the parameter $V_1$ of the intersite Coulomb
interaction, only the three phases corresponding to the
$d_{xy}$-wave, $p$-wave, and $s$-wave types of symmetry of the
superconducting order parameter are stabilized. Note that in the
range of high electron densities and for $0.25<V_1/|t_1|<0.5$, the
Kohn-Luttinger polarization effects lead to the occurrence of the
unconventional $s$-wave pairing~\cite{Kagan13,Kagan14a,Kagan15b}.

Despite their parametric smallness, the second-order effects in
$V$ represent the decisive contribution to superconductivity in
the Shubin-Vonsovsky model. To answer the question why the
first-order contributions in $V$ do not suppress the second-order
contributions, it is necessary to compare the different partial
contributions to the total effective interaction. The histogram in
Fig.~\ref{histogram}b reflects the results for a point of the
phase diagram in which the superconducting phase with the $s$-wave
symmetry of the order parameter corresponds to the ground state.
Matrix elements of the effective interaction $\{U_{eff}\}_{mn}$
calculated for small $m$ and $n$ are presented here in the
histogram. The values of the matrix elements for $n,m> 2$ are not
given because of their smallness. One can see from the histogram
that, for the chosen parameters, the first- and the second-order
contributions $U_{eff}^{(I)}$ and $U^{(U^2)}_{eff}$, respectively,
give only positive values of the matrix elements, and thus
correspond to repulsion. It means that an account of only these
processes would not lead to the $s$-wave superconducting pairing.
Similarly, the second-order contributions $U^{(UV)}_{eff}$ also
would not give rise to superconductivity, and only the
second-order contributions $U^{(V^2)}_{eff}$ provide the negative
values of the matrix elements $\{U_{eff}\}_{mn}$ (and as a result,
the negative eigenvalues of $\lambda$) leading to the realization
of the superconducting $s$-wave pairing.

Thus, the long-range Coulomb repulsion in the lattice models
usually contribute only to the certain pairing channels and does
not affect the other channels. At the same time, the polarization
contributions described by the Kohn-Luttinger diagrams possess the
components in all the channels, and more than one of them usually
plays in favor of attraction. In such a situation, the long-range
Coulomb repulsion probably either does not affect at all the main
component of the effective interaction, which leads to pairing, or
it suppresses the principal component without affecting the
secondary ones~\cite{Raghu12,Kagan13}.

The same scenario of superconductivity is also observed in the
$p$-wave channel in the phase diagram (Fig.~\ref{histogram}a).
\begin{figure}
\begin{center}
  \includegraphics[width=0.38\textwidth]{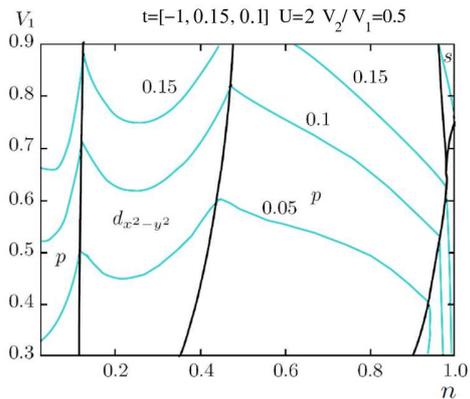}
\caption{Phase diagram of the Shubin-Vonsovsky model calculated
for the parameters $t_2=0.15,~t_3=0.1,~U=2$, and $V_2/V_1=0.5$
(all the parameters are in units of $|t_1|$). Blue curves are the
lines of constant value of $|\lambda|$.}
\label{SV_t2}       % Give a unique label
\end{center}
\end{figure}

In~\cite{Kagan13}, the effect of the distant electron hoppings
$t_2$ and $t_3$ on the superconducting phase diagram was analyzed.
It is known that an account for these hoppings can considerably
modify the density of states (DOS) and shift the Van Hove
singularity (VHS) away from the half-filling to the region of the
lower (or higher) electron densities. Figure~\ref{SV_t2} shows the
modification of the phase diagram of the Shubin-Vonsovsky model,
which is observed upon an increase in $U$. It can be seen that in
the range of low electron densities, as well as in the range of
densities close to the VHS, the $d_{x^2-y^2}$-wave pairing is
achieved with quite large values of $|\lambda|\sim0.1-0.2$. This
result is important for analyzing the possibility of the
realization of the Kohn-Luttinger mechanism in high-T$_c$
superconductors. It should be noted that for $|\lambda|\sim0.2$,
the superconducting transition temperatures can reach the values
$T^{d_{x^2-y^2}}_c\sim 100$ K which are quite reasonable for
cuprates.

\section{The Kohn-Luttinger Superconductivity in Idealized Monolayer and Bilayer Graphene }
\label{sec4}

At the present time, the possible development of superconductivity
in the framework of the Kohn-Luttinger mechanism in graphene under
appropriate experimental conditions is widely discussed. Despite
the fact that intrinsic superconductivity so far has not been
observed in graphene, the stability of the Kohn-Luttinger
superconducting phase has been investigated and the symmetry of
the order parameter on the hexagonal lattice was identified. It
was found~\cite{Nandkishore12} that chiral
superconductivity~\cite{Volovik92} with the $d+id$-wave symmetry
of the order parameter prevails in a large domain near the VHS in
the DOS~\cite{Black14}. The competition between the
superconducting phases with different symmetry types in the wide
electron density range $1<n\leq n_{VH}$, where $n_{VH}$ is the Van
Hove filling, in graphene monolayer was studied in
papers~\cite{Kagan14a,Nandkishore14}. It was demonstrated that at
intermediate electron densities the Coulomb interaction of
electrons located on the nearest carbon atoms facilitates
implementation of superconductivity with the $f$-wave symmetry of
the order parameter, while at approaching the VHS, the
superconducting $d+id$-wave pairing
evolves~\cite{Kagan14a,Nandkishore14}.
\begin{figure}
\begin{center}
  \includegraphics[width=0.4\textwidth]{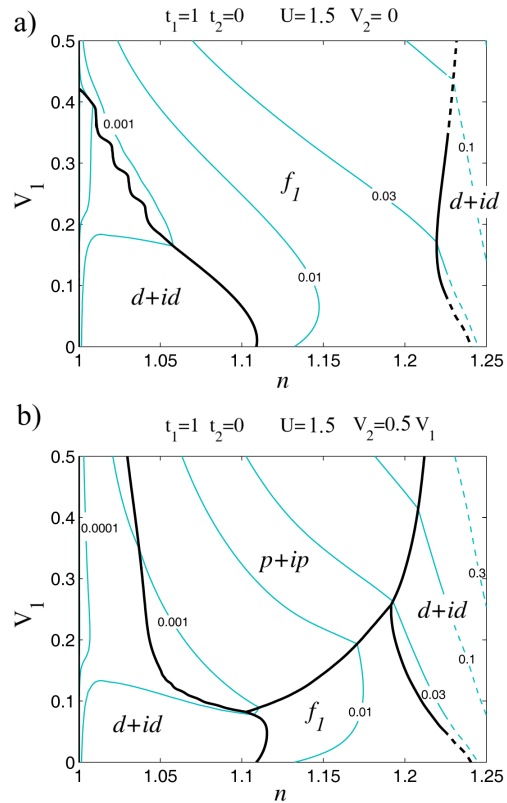}
\caption{(Color online) Phase diagram of the superconducting state
of the graphene monolayer at $U=1.5|t_1|$ for (a) $V_2=0$ and (b)
$V_2=0.5V_1$. Blue curves show the lines of the constant values of
$|\lambda|$.} \label{PD}
\end{center}
\end{figure}

Using the Shubin-Vonsovsky model in the Born weak-coupling
approximation, we investigated the role of the Coulomb repulsion
of electrons located at next- nearest neighboring carbon atoms for
the development of the Kohn-Luttinger superconductivity in an
idealized graphene monolayer disregarding the effect of the Van
der Waals potential of the substrate and both magnetic and
non-magnetic impurities. Figure~\ref{PD}a shows the calculated
phase diagram of the Kohn-Luttinger superconducting state in
graphene monolayer as a function of the carrier concentration $n$
and $V_1$ for the set of parameters $U=1.5|t_1|$, and $V_2 = 0$.
It can be seen that the phase diagram consists of three regions.
At low electron densities $n$, the ground state of the system
corresponds to the chiral superconductivity with the $d+id$-wave
symmetry of the order parameter~\cite{Black14}. At the
intermediate electron densities, the superconducting $f$-wave
pairing is implemented. At the large values of $n$, the domain of
the superconducting $d+id$-wave pairing
occurs~\cite{Nandkishore12}. With the increase of the parameter
$V_1$ of the intersite Coulomb interaction, in the region of small
values of $n$, the $d+id$-wave pairing is suppressed and the
pairing with the $f$-wave symmetry of the order parameter is
implemented. Thin blue lines in Fig.~\ref{PD} are the lines of the
equal values of the effective coupling constant $|\lambda|$. It
can be seen that in this case in the vicinity of $n_{VH}$ the
effective coupling constant reaches the values of $|\lambda|=0.1$.
\begin{figure}
\begin{center}
  \includegraphics[width=0.42\textwidth]{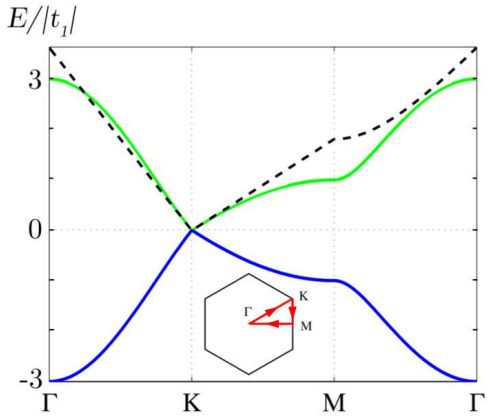}
\caption{(Color online) Energy spectra of graphene monolayer (blue
and green solid lines) and energy spectra obtained in the
framework of the Dirac approximation (black dashed line). Subplot
depicts the path around the Brillouin zone.} \label{Dirac}
\end{center}
\end{figure}

Let us consider the modification of the phase diagram for graphene
monolayer with respect to the Coulomb interaction $V_2$ between
the electrons located at the next-nearest carbon atoms. It can be
seen in Fig.~\ref{PD}b for the fixed ratio between the parameters
of the long-range Coulomb interactions $V_2=0.5V_1$ that when
$V_2$ is taken into account, the phase diagram changes
qualitatively. These changes involve the suppression of a large
region of the $f$-wave pairing at the intermediate electron
densities and the implementation of the chiral superconducting
$p+ip$-wave pairing. In addition, when $V_2$ is taken into
account, the effective coupling constant increases to the values
of $|\lambda|=0.3$. Consequently, it leads to a significant
increase in $T_c$ in idealized doped graphene. Note that here we
do not analyze the account for the electron hoppings to the
next-nearest carbon atoms $t_2$, since an account for these
hoppings for graphene monolayer does not significantly modify the
DOS in the carrier concentration regions between the Dirac point
and both VHS points $n_{VH}$~\cite{Kagan14a}.

It should be noted also that the Kohn-Luttinger superconductivity
in graphene never develops near the Dirac points. The calculations
show that in the vicinity of these points, where the linear
approximation for the energy spectrum of graphene monolayer works
pretty well, the DOS is very low and the effective coupling
constant $|\lambda|<10^{-2}$. The higher values of $|\lambda|$,
which are indicative of the development of the Cooper instability
at reasonable temperatures, arise at the electron densities
$n>1.15$. However, at such densities, the energy spectrum of the
monolayer along the direction $KM$ of the Brillouin zone
(Fig.~\ref{Dirac}) already significantly differs from the Dirac
approximation.

\begin{figure}
\begin{center}
  \includegraphics[width=0.42\textwidth]{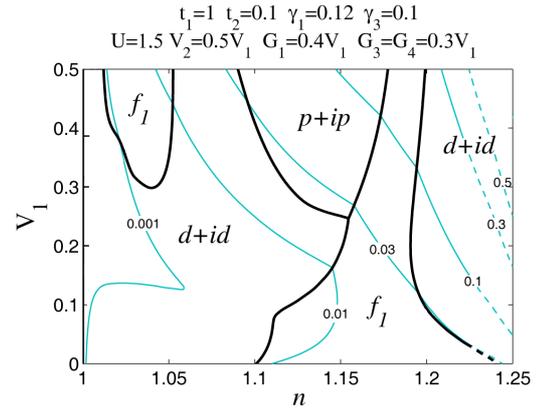}
\caption{Phase diagram of the superconducting state of graphene
bilayer shown as a function of the variables "$n-V_1$" at
$t_2=0.1,\,\gamma_1=0.12,\,\gamma_3=0.1,\,U=1.5,\,V_2=0.5V_1,\,G_1=0.4V_1,\,G_3=G_4=0.3V_1$
(all the parameters are in units of $|t_1|$). $G_1$, $G_3$ and
$G_4$ are the parameters of the interlayer Coulomb repulsion,
$\gamma_1$ and $\gamma_3$ are the interlayer hoppings. Blue curves
are the lines of the constant values of $|\lambda|$.}\label{PD_t2}
\end{center}
\end{figure}

Since the electronic properties of graphene depend on the number
of carbon layers~\cite{Guinea06}, we analyzed the possibility of
implementation of the Kohn-Luttinger superconductivity and
constructed the phase diagram (Fig.~\ref{PD_t2}) in idealized
graphene bilayer~\cite{Kagan15a}.
The calculation shows that the separate increase of the parameters
of the interlayer Coulomb repulsion $G_3$ and $G_4$ suppresses the
$d+id$-wave pairing and, at the same time, broadens the $f$-wave
pairing region at small electron densities. The superconducting
$d+id$-wave phase is suppressed the most effectively by enhancing
the parameter $G_4$ of the interlayer Coulomb interaction. When
the interactions $G_3$ and $G_4$ are simultaneously taken into
account, then along with the intensive suppression of the
superconducting $d+id$-wave pairing at small electron densities
and the implementation of the superconductivity with the $f$-wave
symmetry of the order parameter, the growth of the absolute values
of effective coupling constant $\lambda$ is also observed.

\section{The Kohn-Luttinger superconductivity in real graphene}
\label{sec5}

Our calculation showed that the Kohn-Luttinger mechanism can lead
to the superconducting transition temperatures $T_c\sim
10\div20~\textrm{K}$ in an idealized graphene monolayer and
$T_c\sim 20\div40~\textrm{K}$ in an idealized graphene bilayer.
Contrary to these rather optimistic estimations, in real graphene,
superconductivity has not been found yet. This material is only
close to superconductivity.

The reason for that is probably connected with the effects of
structural disorder and the presence of the nonmagnetic impurities
in real graphene (or graphite). Note that for exotic ($p$-, $d$-
and $f$-wave) as well as for $s$-wave superconducting pairing with
nodal points on the 2D lattice ($\Delta_s(\phi)\sim\cos 6n\phi$,
$\Delta_{s_{ext}}(\phi)\sim\sin 6n\phi$, $n\geq1$) the Anderson
theorem~\cite{Anderson59} for nonmagnetic impurities is violated
and anomalous superconductivity can be suppressed for $\gamma\geq
k_BT_c^{clean}$~\cite{Abrikosov61,Abrikosov63,Posazhennikova96},
where $\gamma = \hbar/(2\tau)$ is an electron damping due to
scattering on impurities. In another words, $k_BT_c^{clean} \leq
\hbar p_F/ml$, where the Fermi-momentum $p_F$ is connected with a
2D electron density $n_{2D}$ in graphene ($p_F\sim\hbar\sqrt{2\pi
n_{2D}}$, if we assume the circular Fermi-surface for simplicity).
At the same time, $l$ is an effective mean-free path extracted
both from transport and Hall measurements and thus taking into
account both the effects of the structural disorder and the
presence of the nonmagnetic impurities.

To our best knowledge, the record experimental parameters
available nowadays correspond to $\textrm{max}(l)\sim 2\cdot10^3
\textrm{\AA}$ and
$\textrm{max}(n_{2D})\sim10^{13}\,\textrm{cm}^{-2}$ in real
graphene monolayers. The maximal values of the mean-free path
today still correspond to the moderately clean case. The
corresponding amount of disorder for the high density $n_{2D}\sim
10^{13}\,\textrm{cm}^{-2}$ is sufficient to suppress totally
anomalous $T_c$ of the order of $10~\textrm{K}$. This experimental
challenge for the discovery of superconductivity in real graphene
is to further increase the 2D electron density or to prepare the
ultraclean graphene monolayer or bilayer. Another possibility for
getting closer to realization of superconductivity in real
graphene is to perform the experiments on quasi 1D epitaxial
graphene nanoribbons~\cite{Baringhaus14}. In this case, however,
the competition between superconductivity and the Peierls-type of
instabilities is highly possible, at least on the level of
theoretical considerations in the framework of parquet
diagrammatic approximation~\cite{Bychkov66}.

\begin{acknowledgements}
The authors are grateful to V.\,V. Val'kov, M.\,V. Feigel'man and
A.\,Ya. Tzalenchuk for valuable remarks. This work is supported by
the Russian Foundation for Basic Research (nos. 14-02-00058 and
14-02-31237). One of the authors (M.\,Yu.\,K.) gratefully
acknowledges support from the Basic Research Program of the
National Research University Higher School of Economics. Another
one (M.\,M.\,K.) thanks the scholarship SP-1361.2015.1 of the
President of Russia and the Dynasty foundation.
\end{acknowledgements}


\begin{thebibliography}{}
\bibitem{Kohn65}
Kohn, W., Luttinger, J.\,M.: Phys. Rev. Lett. \textbf{15}, 524
(1965)

\bibitem{Fay68}
Fay, D., Layzer, A.: Phys. Rev. Lett. {\bf 20}, 187 (1968)

\bibitem{Kagan88}
Kagan, M.\,Yu., Chubukov, A.\,V.: JETP Lett. \textbf{47}, 614
(1988)

\bibitem{Baranov92b}
Baranov, M.\,A., Chubukov, A.\,V., Kagan, M.\,Yu.: Int. J. Mod.
Phys. B \textbf{6}, 2471 (1992)

\bibitem{Kagan15b}
Kagan, M. Yu., Mitskan, V. A., Korovushkin, M. M.: Phys. Usp.
\textbf{58} (8) (2015)

\bibitem{Hubbard63}
Hubbard, J.\,C.: Proc. R. Soc. London A {\bf 276}, 238 (1963)

\bibitem{Baranov92a}
Baranov, M. A., Kagan, M. Yu.: Z. Phys. B: Condens. Matter
\textbf{86}, 237 (1992)

\bibitem{Hlubina99}
Hlubina, R.: Phys. Rev. B \textbf{59}, 9600 (1999)

\bibitem{Raghu12}
Raghu, S., Berg, E., Chubukov, A.\,V., Kivelson, S.\,A.: Phys.
Rev. B {\bf 85}, 024516 (2012)

\bibitem{Kagan13}
Kagan, M. Yu., Val'kov, V. V., Mitskan, V. A., Korovushkin, M. M.:
JETP Lett. \textbf{97}, 226 (2013); JETP \textbf{117}, 728 (2013)

\bibitem{Raghu10}
Raghu, S., Kivelson, S.\,A., Scalapino, D.\,J.: Phys. Rev. B
\textbf{81}, 224505 (2010)

\bibitem{Alexandrov11}
Alexandrov, A.\,S, Kabanov, V.\,V.: Phys. Rev. Lett. \textbf{106},
136403 (2011)

\bibitem{Kagan11}
Kagan, M.\,Yu., Efremov, D.\,V., Marienko, M.\,S., Val'kov,
V.\,V.: JETP Lett. \textbf{93}, 720 (2011)

\bibitem{Shubin34}
Shubin, S., Vonsowsky, S.: Proc. Roy. Soc. A {\bf 145}, 159 (1934)

\bibitem{Nandkishore12}
Nandkishore, R., Levitov, L.\,S., Chubukov, A.\,V.: Nature Phys.
\textbf{8}, 158 (2012)

\bibitem{Volovik92}
Volovik, G.\,E.: \emph{Exotic properties of superfluid $^3$He},
World Scientific, Singapore, 1992.

\bibitem{Black14}
Black-Schaffer, A.\,M., Honerkamp, C.: J. Phys.: Condens. Matter
\textbf{26}, 423201 (2014)

\bibitem{Kagan14a}
Kagan, M. Yu., Val'kov, V. V., Mitskan, V. A., Korovushkin, M. M.:
Solid State Commun. \textbf{188}, 61 (2014)

\bibitem{Nandkishore14}
Nandkishore, R., Thomale, R., Chubukov, A.\,V.: Phys. Rev. B
\textbf{89}, 144501 (2014)

\bibitem{Guinea06}
Guinea, F., Castro Neto, A.\,H., Peres, N.\,M.\,R.: Phys. Rev. B
\textbf{73} 245426 (2006)

\bibitem{Kagan15a}
Kagan, M. Yu., Mitskan, V. A., Korovushkin, M. M.: Eur. Phys. J. B
\textbf{88}, 157 (2015)

\bibitem{Anderson59}
Anderson, P.\,W.: J. Phys. Chem. Solids \textbf{11}, 26 (1959)

\bibitem{Abrikosov61}
Abrikosov, A.\,A., Gor'kov, L.\,P.: Sov. Phys. JETP \textbf{12},
337 (1961)

\bibitem{Abrikosov63}
Abrikosov, A.\,A., Gor'kov, L.\,P., Dzyaloshinski, I.\,E.:
\emph{Quantum Field Theory in Statistical Physics}, Prentice Hall,
Englewood Cliffs, New Jersey, 1963

\bibitem{Posazhennikova96}
Posazhennikova, A.\,I., Sadovskii, M.\,V.: JETP Lett. {\bf 63},
358 (1996)

\bibitem{Baringhaus14}
Baringhaus, J., Ruan, M., Edler, F., et al.: Nature {\bf 506}, 349
(2014)

\bibitem{Bychkov66}
Bychkov, Yu.\,A., Gor'kov, L.\,P., Dzyaloshinskii, I.\,E.: Sov.
Phys. JETP \textbf{23}, 489 (1966)












\end{thebibliography}
\end{document}